\documentclass[%
 aip,
 amsmath,amssymb,
 reprint,%
]{revtex4-1}
\draft
\usepackage{graphicx}  
\usepackage{dcolumn}
\usepackage{amsmath}
\usepackage{bm}        
\usepackage{amssymb}   

\usepackage[utf8]{inputenc}
\usepackage[T1]{fontenc}
\usepackage{mathptmx}

\usepackage{float}
\usepackage{color}

\newcommand{\ntet}{n_\mathrm{tet}}
\newcommand{\ntetav}{\left< n_\mathrm{tet} \right>}
\newcommand{\sav}{\left< S_\mathrm{2}\right>}
\newcommand{\gmax}{ \mathrm{\gamma}_\mathrm{max}}

\begin{document}
\title{{Correlation between plastic rearrangements and local structure in a cyclically driven glass}}

\author{Saheli Mitra}
\affiliation{Université Paris-Saclay, CNRS, Laboratoire de Physique des Solides, 91405, Orsay, France.}

\author{Susana Mar\'in-Aguilar}
\email{s.marinaguilar@uu.nl}
\affiliation{Université Paris-Saclay, CNRS, Laboratoire de Physique des Solides, 91405, Orsay, France.}
\affiliation{Soft Condensed Matter, Debye Institute for Nanomaterials Science, Department of Physics, Utrecht University, Princetonplein 1, 3584 CC Utrecht, The Netherlands.}

\author{Srikanth Sastry}\affiliation{Jawaharlal
  Nehru Center for Advanced Scientific Research, Jakkur Campus, Bengaluru 560064, India.}

\author{Frank Smallenburg}
\email{frank.smallenburg@universite-paris-saclay.fr}
\affiliation{Université Paris-Saclay, CNRS, Laboratoire de Physique des Solides, 91405, Orsay, France.}

\author{Giuseppe Foffi}
\email{giuseppe.foffi@universite-paris-saclay.fr}
\affiliation{Université Paris-Saclay, CNRS, Laboratoire de Physique des Solides, 91405, Orsay, France.}

\date{\today}
\begin{abstract}

{The correlation between local structure and the propensity for structural rearrangements has been widely investigated in glass forming liquids and glasses.}\\
In this paper we use the excess two-body entropy $S_2$ and tetrahedrality $\ntet$ as the per-particle local structural order parameters to explore such correlations in a three-dimensional model glass subjected to cyclic shear deformation. We first show that for both liquid configurations and the corresponding inherent structures, local ordering increases upon lowering temperature, signaled by a decrease in the two-body entropy and an increase in tetrahedrality. When the inherent structures, or glasses, are periodically sheared athermally, they eventually reach {\it absorbing} states for small shear amplitudes, which do not change from one cycle to the next. Large strain amplitudes result in the the formation of shear bands, within which particle motion is diffusive. 
We show that in the steady state, there is a clear difference in the local structural environment of particles that will be part of plastic rearrangements during the next shear cycle and that of particles that are immobile. In particular,  particles with higher $S_2$ and lower $\ntet$ are more likely to go through rearrangements irrespective of the average energies of the configurations and strain amplitude. For high shear, we find very distinctive local order outside the mobile shear band region, where almost $30\%$ of the particles are involved in icosahedral clusters, contrasting strongly with the fraction of $<5\%$ found inside the shear band.
\end{abstract}
\maketitle

\section{Introduction}

The mechanism involved in how a liquid loses its fluidity upon decreasing temperature or increasing density has been {the subject of intense research investigations} for several decades now~\cite{berthier2011theoretical}. 
One interesting feature that emerges in liquids close to the glass transition is dynamical heterogeneity, where over time scales corresponding to the $\alpha$-relaxation time, some regions of the systems have significantly higher mobility than others.
A growing body of literature  now supports the existence of a correlation between this heterogeneity and the local structure~\cite{ royall2015role, schoenholz2016structural,tanaka2019revealing,boattini2020autonomously,Paret_2020}. One of the pioneering ideas in analysing the local structure on glasses was proposed  by Charles Frank in 1952 indicating the possible prevalence of icosahedral clusters in the glassy regime \cite{frank1952supercooling}. Later, many studies corroborated the presence of locally favored structures in the glassy regime related to the slowdown of dynamics, with  icosahedral clusters being the focus of attention in several studies \cite{jonsson1988icosahedral,kondo1991icosahedral,steinhardt1981icosahedral, jonsson1988icosahedral, hirata2013geometric, ding2014full, marin2019slowing, marin2021monodisperse}. Along similar lines, there have been proposals of simpler tetrahedron-based order parameters capable of capturing the changes in local structure and dynamics\cite{tong2018revealing,xia2015structural,anikeenko2007polytetrahedral,charbonneau2012geometrical, hirata2013geometric,marin2020tetrahedrality}. 
In particular, a recent development was the introduction of the concept of tetrahedrality of the local structure $\ntet$, which measures the number of tetrahedral clusters each particle is involved in~\cite{marin2020tetrahedrality}, based on the notion that, at least for some glass formers, most of the locally favoured structures (including icosahedral) can be decomposed into tetrahedra. It was found that the particles with higher values of $\ntet$ are strongly correlated with slower dynamics. \\
These studies are based strictly on local structures whose ability to fully predict glassy dynamics has been questioned~\cite{Berthier_2007}. In general, while a full understanding of the dynamics may require information beyond the static local environment, local structures represent an important starting point as testified by the success of machine learning based approaches that can be trained to predict the local mobility of different regions in a glassy fluid ~\cite{schoenholz2016structural,Bapst_2020, boattini2021averaging}. Moreover, the relevance of icosahedral and polytetrahedral local structure has been recently highlighted  by novel unsupervised machine learning approaches, such as community inference~\cite{Paret_2020} and neural-network-based auto-encoders\cite{boattini2020autonomously}.\\
Apart from the heterogeneity in supercooled liquids, local structure can play an important role in the  plastic rearrangements occurring in a sheared solid.
When shearing a crystalline solid, rearrangement events are normally expected to occur in the vicinity of crystal defects. This idea can be extended to sheared glasses, where certain spots in the system may be ``weaker'' than others, and hence more prone to plastic rearrangement. These ``soft spots''\cite{widmer2008irreversible,chen2011measurement,manning2011vibrational} or ``shear transformation zones''\cite{falk1998dsqmin,argon1979plastic} can be seen as analogues of crystal defects in glassy materials. 
In recent years, a variety of approaches have been explored to predict such ``soft spots'', including e.g. machine learning \cite{cubuk2015identifying}, local vibrational modes \cite{manning2011vibrational},  local yield stress \cite{patinet2016connecting}.
A summary of the efficiency of the different methods is reported in a recent paper by Richard et al. \cite{richard2020predicting},  in which the authors compare the performance of a variety of methods  to  predict  plasticity due to uniform shear deformation.  For a  $2d$ glass former,  they demonstrated that rearrangements are indeed deeply encoded in the structure~ \cite{richard2020predicting, bhaumik2021yielding}. At the experimental level,  soft colloidal glasses under thermal cycling display high correlation between local rearrangements and local two-body entropy\cite{yang2016structures}. \\
In the aforementioned works, the role of local structure in amorphous systems undergoing \textit{cyclic} mechanical deformation has not been studied in detail.  Cyclic shear deformation is a commonly  used technique to test mechanical properties of materials\cite{rogers2011sequence,koumakis2013complex,gibaud2016multiple}, memory effects \cite{keim2011generic,fiocco2014encoding}, self organization\cite{royer2015precisely}, and annealing of glass \cite{das2018annealing}.  When a glass is deformed in the quasi-static limit at zero temperature \cite{maloney2006amorphous}, compared to uniform shearing, the cyclic shear displays  a sharp yielding transition from an absorbing to a diffusive state at a critical strain amplitude $\gamma_y$ \cite{leishangthem2017yielding}. Below this critical threshold  $\gmax < \gamma_y$, after several training cycles of deformation, the system reaches a steady  state. In this state, the plastic rearrangements during a deformation cycle are fully reversible\cite{fiocco2014encoding,Adhikari2018,priezjev2016reversible} and localized. Above yielding,  rearrangements are connected to large displacements and are irreversible. As a consequence plastic events become  spatially organized, leading to the appearance of a  shear band \cite{parmar2019strain, VasishtPRE2020b}. \\
In this paper we use a binary repulsive $3d$ glass former to explore the structural properties of  particles undergoing plastic rearrangements due to cyclic shear deformation. We analyze how the local structural order of the system is correlated with these rearrangements. In particular, we compute the per-particle  two-body excess entropy $S_2$ and tetrahedrality $\ntet$, the latter not having been  explored yet in the context of plastic rearrangements. Our analysis  focuses on  the steady state and we look for structural differences between the local environments for {subsets of} particles that exhibit the largest and {the smallest} displacements in successive cycles. \\
The paper is organized as follows:
In section II, we discuss the glass former model in detail and we describe the simulation methods. In section III we introduce the structural descriptors. Thereafter,
in section IV, we present our results in the following order. We first report (\textit{i}) the variation of mean local ordering in liquid and inherent structures (IS)  using the $\ntetav$ and $\sav$.
(\textit{ii}) We correlate the liquid structures and the corresponding IS at different temperatures. We then sample a high and a low temperature glass and shear them athermally for many deformation cycles with strain amplitudes $\gmax$. (\textit{iii}) In the steady state we compute the mean values of our structural order parameters as $\gmax$ varies in a range across the yielding amplitude $\gamma_y$. (\textit{iv}) From one cycle to the next we classify \textit{mobile} and \textit{immobile} particles with larger and smaller rearrangements respectively by computing the non-affine displacements  $D^2_{min}$ \cite{falk1998dsqmin}. (\textit{v}) In terms of $\ntet$ and $S_2$ we examine whether there is a difference between the mean local order of the mobile and immobile particles. (\textit{vi}) Later, we use the topological cluster classification algorithm (TCC)~\cite{malins2013identification} to point out the different cluster association between the two classes. (\textit{vii}) Finally, we explore the structure inside and outside the shear band for the sheared glasses above the yielding transition.\\
Our results show that there is a clear structural difference in local arrangements of the particles involved in the largest and smallest plastic rearrangements. 
Similarly, above yielding, we find that particles inside and outside the shear band have strongly different local environments.
\begin{center}
 \begin{figure*}[!htb]
   \includegraphics[width=1\linewidth]{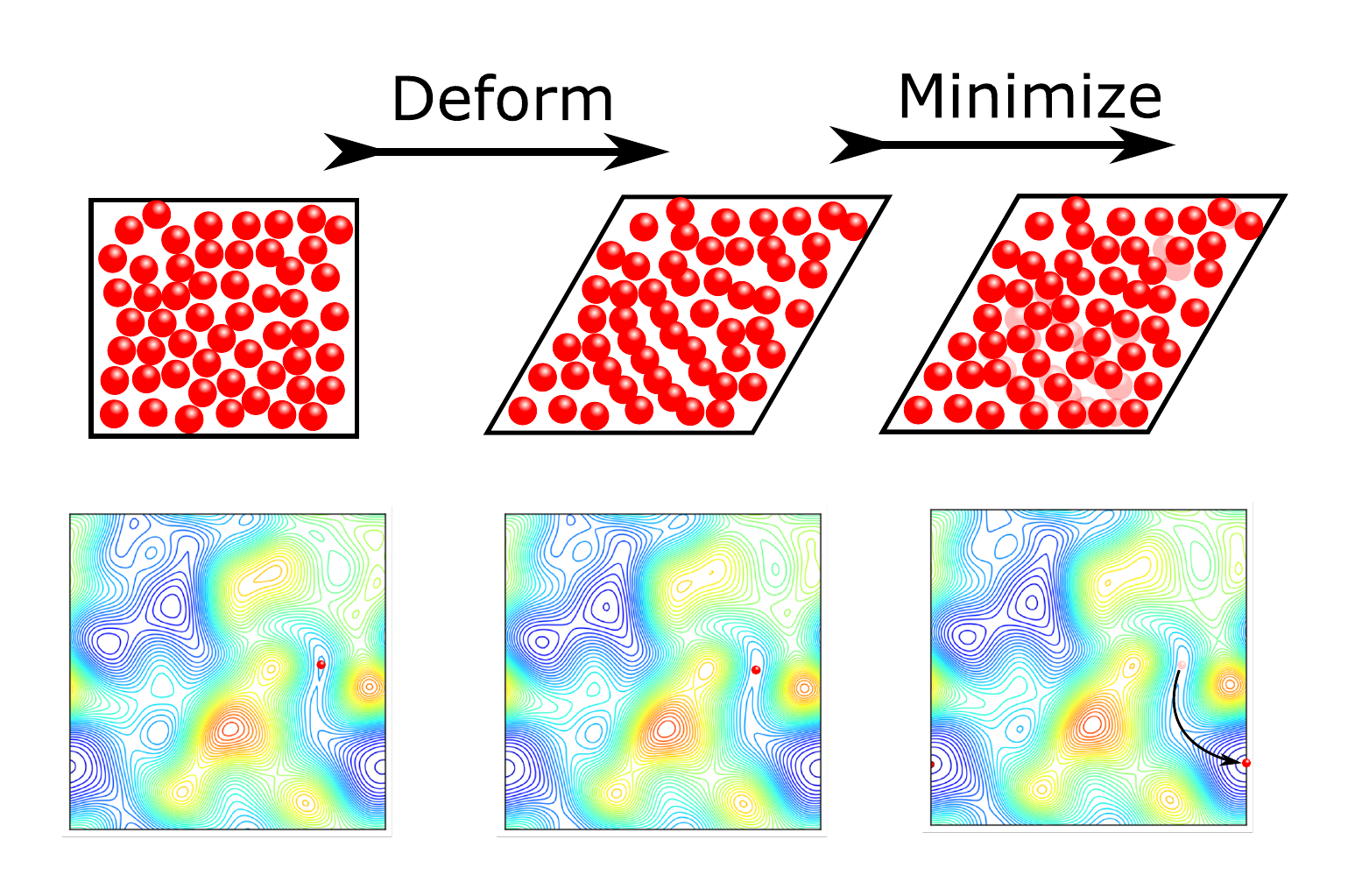}
    \caption{ Schematic representation of an Athermal Quasi Static  (AQS) deformation move. In the top row we present a cartoon of the configuration and in the bottom one a simplified potential energy landscape. An initial configuration in its energy minima (left column) is deformed with small strain steps (central column). In this specific case, the system is not at a minimum anymore and the following minimization leads it into a new minima that corresponds to a new configuration (right column). }
    \label{fig:0}
 \end{figure*}
\end{center}
\section{Methods}
\subsection{Glass-former model and Simulation}
As a model glass system, we explore the behavior of purely repulsive Wahnstr\"om (WH) model \cite{wahnstrom1991molecular}. It consists of a $50:50$ mixture of particles {with additive diameters} interacting through a Lennard-Jones (LJ) potential,
\begin{equation}
    V_{\alpha\beta}(r)= \begin{cases}
    4 \epsilon_{\alpha\beta}\left[ \left( \frac{\sigma_{\alpha\beta}}{r} \right)^{12} 
                        - \left( \frac{\sigma_{\alpha\beta}}{r} \right)^{6} \right] - V_{r_c}, &
                        r < r_{c}~~~\\
                        0, & r \geq r_{c}
                        \end{cases}
\end{equation}
where $\alpha$ and $\beta$ denote the type of particle (A or B), the cut-off distance is defined as $r_{c}=2^{\frac{1}{6}}\sigma_{\alpha\beta}$ and $V_{r_c}$ is the value of the LJ potential evaluated at $r_c$. With this choice of the cut-off, only the repulsive part of the potential is retained. The potential parameters are defined with respect to type $A$: $\sigma_{BB}/\sigma_{AA}=1.2$, $\sigma_{AB}/\sigma_{AA}=1.1$ and $\epsilon_{AA}=\epsilon_{BB}=\epsilon_{AB}$. We fix the total packing fraction to $\phi=\frac{\pi}{6V}(N_A \sigma_{AA}^3+N_B\sigma_{BB}^3) = 0.58$, where $V$ is the simulation box volume and the total number of particles $N=N_A + N_B = 64000$.

We perform molecular dynamics simulations using LAMMPS \cite{plimpton1993fast}  with a fixed time step size of $dt=0.005 \tau$, where $\tau=\sqrt{m \sigma_{AA}^2/\epsilon_{AA}}$ is the time unit, with $m$ the particle mass. We use reduced LJ units where $\sigma_{AA}$ is the unit for distance and $\epsilon_{AA}$ the energy unit. We prepare a set of equilibrated configurations at temperature $T$ by  running $NVT$ molecular dynamics for $2 \times 10^5$ (high $T$) - $2 \times 10^6$ (low $T$) time steps followed by constant energy ($NVE$) relaxation. From these equilibrated samples, we obtain the corresponding inherent structures (IS) by minimizing the potential energy, using conjugate gradient minimization, with a tolerance $10^{-16}$. {The IS configurations are the "glass" configurations which we subject to shear deformation, which we associate with the temperature of the liquid simulations from which they are generated.}
The range of $T\in[0.7,2.0]$ for this WH system is sufficient to show the dynamical slowdown from high to low temperature as expected in a typical glass-former. 

Next, to study the plastic re-arrangements in glasses, we focus on one high temperature ($T=1.5$) and one low temperature ($T=0.7$) glass, both quenched to the IS and sheared using the athermal quasi static shear  (AQS) protocol \cite{maloney2004subextensive,maloney2006amorphous}. In AQS, the $xz$ plane of a chosen IS configuration is sheared in small strain steps $d\gamma=0.0002$ in one direction and each deformation step is followed by an energy minimization. 
{A schematic representation of an AQS step is presented in Fig.~\ref{fig:0}. 
This is continued until the strain values reach a specified amplitude $\gamma_{max}$ in either direction, at which the strain direction is reversed.}
{Therefore, one complete cycle of deformation has the following sequence:}
\begin{align}
 \gamma=0 \rightarrow \gamma_{max} \rightarrow 0 \rightarrow -\gamma_{max}\rightarrow 0.
\end{align}
Hence, each cycle consists of $4*\gmax/d\gamma$  steps. 

{For each value of the}
parameter $\gmax$, we deform the system for $100-600$ cycles until a steady state is reached. The energy of stroboscopic configurations, sampled at the end of each cycle with $\gamma=0$, will depend on $\gmax$.  At $\gmax$ below the yielding transition, $\gmax <\gamma_y$ where $\gamma_y$ is the yielding amplitud of the system, {the energies attain a constant value in the steady state.}
For ($\gmax > \gamma_y$), the energies fluctuate around a mean value in the steady state. In our WH system, $\gamma_y \approx 0.06$. For higher amplitudes than $\gamma_y$ the system yields, becoming diffusive.  

\subsection{Local structure observables}
\subsubsection{Local entropy}

To characterize local structure at the particle level, it is useful to first consider a structural descriptor that is known to correlate with dynamical slowdown below the onset temperature. For this purpose, we consider the local entropy $S_2$, defined in terms of pairwise structural correlations.
In general, the entropy $S$ can be expanded in terms of multi-particle correlation functions as $S=S_1+S_2+S_3+ ..$, where $S_1$ is the ideal gas value, $S_2$ is the two-body excess entropy which can be calculated in terms of the pair distribution function, $S_3$ involves three-body correlations,  {\it etc.} \cite{wallace1987role}.  The two-body term  $S_2$\cite{dyre2018perspective,rosenfeld1977relation,rosenfeld1999quasi} has been widely used in the context of transport coefficients and plasticity. In general, it measures the loss of entropy due to positional correlations with a lower value of $S_2$ corresponding to a more ordered structure. Even though $S_2$ has been introduced long ago ~\cite{Nettleton_1958,baranyai1989direct}, it is a useful concept that continues to be widely used~\cite{truskett2000towards, Tanaka_2010,anikeenko2007polytetrahedral, Hallett2018}. As a matter of fact, recently, scaling relations have been reported between $S_2$ and diffusivity in glassy systems  \cite{bell2020excess} and relaxation rates in cyclic sheared systems \cite{galloway2020scaling}. 
In order to look at this quantity on a single-particle level, the two-body entropy of particle $i$ can be defined as:
\begin{align}
    S^i_2 = -2\pi \rho k_B \int_0^\infty [g_m^i(r)\log\left(g_m^i(r)\right)-g_m^i(r)+1]r^2dr,
    \label{eqns2}
\end{align}
where $g_m^i(r)$  is  the \textit{mollified} radial distribution function   \cite{Piaggi2017, Hallett2018},
\begin{align}
   g^i_m(r) = \frac{1}{4 \pi N \rho r^2} \sum_{i \neq j} \frac{1}{2\pi \sigma^2} \exp{[-(r-r_{ij})^2/(2\sigma^2)]}.
\end{align}
with $r_{ij}$ the distance between the $i^{th}$ and $j^{th}$ particle and $\sigma$ a parameter setting the width of the Gaussian kernel. This per-particle radial distribution function is well behaved thanks to the broadening of the neighbor position from a delta peak to a narrow Gaussian. The parameter $\sigma$ is chosen such that the global average of $g_m^i(r)$ is close to the true radial distribution function $g(r)$ and we have a smooth integral to compute $S_2$. For our WH system, we choose  $\sigma=0.09$ and for each particle the integration in Eq. \ref{eqns2} is taken from zero to $r_{max}=5.0$. With this approach we obtain the per particle excess entropy  that we call  simply $S_2$ in the rest of the paper. When averaged over all particles in the system, and over different configurations, we obtain the average two-body entropy $\langle S_2\rangle$.

By its nature, $S_2^i$ provides a measure for the local radial ordering of neighbors around particle $i$: many neighbors at the same absolute distance $r$, as one might find in highly symmetric local environments, will result in a low value of $S_2^i$. Hence, a more negative value of $S_2$ indicates higher local order.  

\subsubsection{Tetrahedrality}

The second descriptor we characterize is the tetrahedrality of the local structure,  $\ntet$, which measures, {for each particle}, the number of  tetrahedra it is involved in~\cite{marin2020tetrahedrality}. As some of the clusters that have been found in supercooled liquids, such as the icosahedral cluster, can be decomposed in a large number of tetrahedron, the tetrahedrality is a good descriptor of the local environment of a particle that correlates strongly with glassy dynamics in hard-sphere-like systems \cite{marin2020tetrahedrality, boattini2020autonomously}. 

In order to quantify the tetrahedrality, we first identify the neighbors of the particles through a modified Voronoi construction identical to the one done in  the Topological Cluster Classification algorithm~\cite{malins2013identification}. We define a tetrahedral cluster as a cluster formed by four particles which are all considered nearest neighbors of each other according to the Voronoi construction. For an individual particle, $\ntet$ is than simply the number of distinct tetrahedral clusters that include this particle. Finally, we can also characterize the global structure of a system through the average of the per-particle tetrahedrality $\ntetav$ \cite{marin2020tetrahedrality}. 
A higher value of $\ntet$ generally is related to more compact and long-lived clusters such as the icosahedral cluster.

\section{Results}

\subsection{Liquid and Inherent Structures}
\begin{figure}[!htb]
\begin{center}
   \includegraphics[width=1\linewidth]{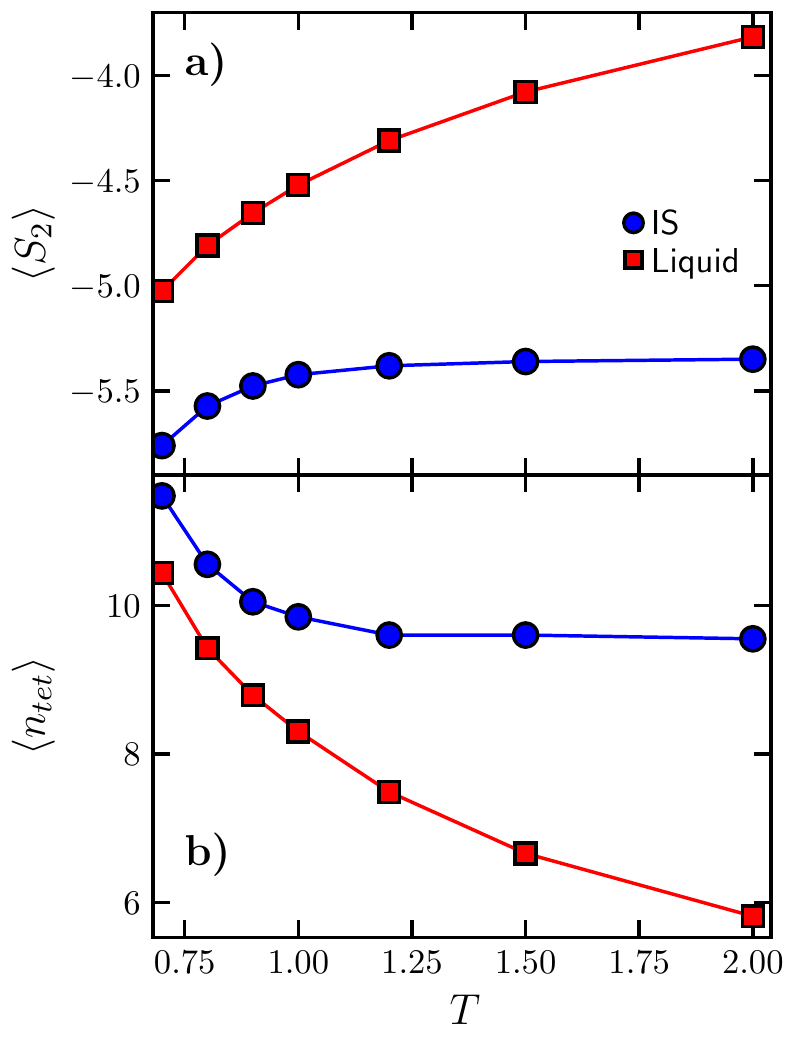}
    \caption{a) Two-body entropy $\sav$ as a function of temperature $T$ for liquid and inherent structure. b) Tetrahedrality $\ntetav$ as a function of $T$ for liquid and inherent structure.}
    \label{fig:1}
\end{center}
\end{figure}

We begin our study by exploring the behavior of both $\sav$ and $\ntetav$ for both liquid and inherent structures at different  temperatures. In Fig.~\ref{fig:1}, we show the dependence of both structural quantities for temperatures between  $T=0.7-2.0$. In the liquid,  upon lowering the temperature, the average local order increases as it can be deduced by  the increase in the value of $\ntetav$ and decrease of $\sav$. As expected, for all temperatures,  the inherent structure presents a higher degree of ordering that translates into a larger $\ntetav$ and smaller $\sav$ in comparison to the liquid. However, for the IS, these quantities  become essentially independent of $T$ at sufficiently high temperatures. This trend is consistent with the previous observation of an increase of the number of locally favored  structures (LFS)~\cite{Hallett2018} when the dynamical slowdown start occurring  and the aforementioned temperature can be identified as the  \textit{onset} temperature at which the dynamics start to be dominated by the potential energy landscape~\cite{Sastry1998}. 
Above this temperature, the dynamics are typical of a high-temperature liquid and only marginally influenced by the energy landscape. As a consequence, in this regime, the system samples several independent minima of the IS   and   $\sav$ and $\ntetav$ are insensitive to temperature variations. This means that the system explores a large range of energy basins with no specific bias toward a particular energetic level. Below the onset temperature, the system starts to explore preferentially basins with a  low energy that decrease with temperature.

\begin{figure}[!htb]
\begin{center}
  \includegraphics[width=1\linewidth]{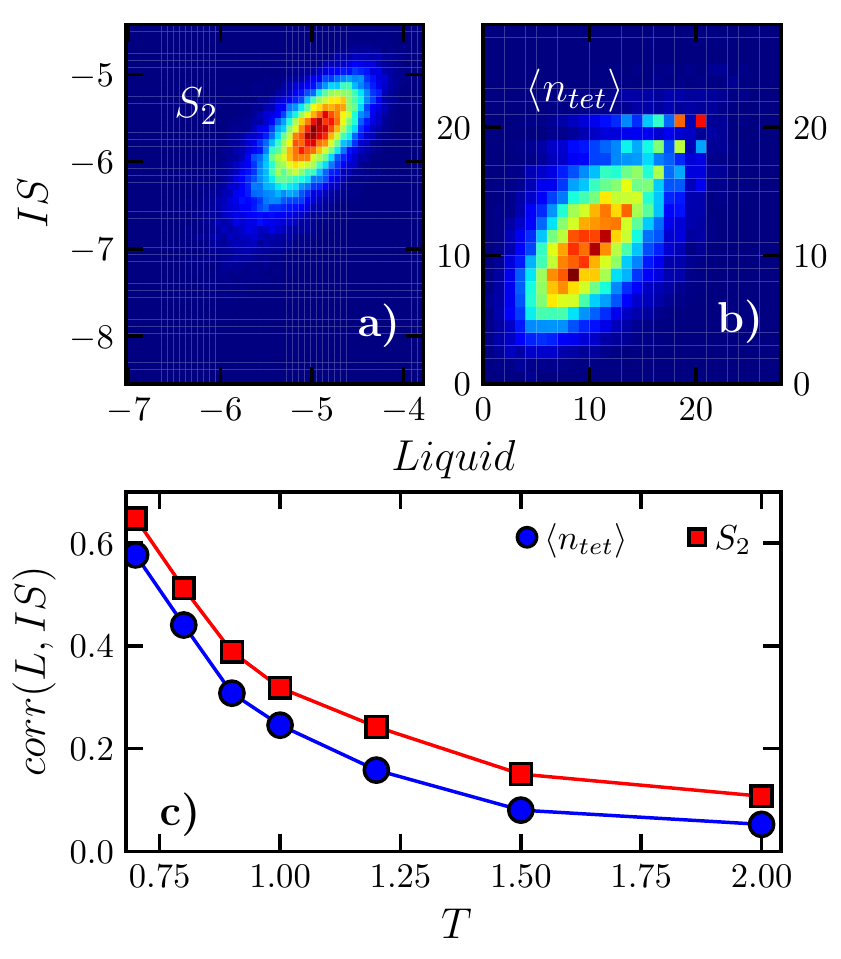}
  \caption{Density plot between inherent structure and liquid of a) $S_2$ and b) $\ntet$ at a temperature $T=0.7$. c) Spearman's correlation between the structural descriptors of liquid and inherent structures as a function of temperature $T$.  }
  \label{fig:2}
\end{center}
\end{figure}

In the low temperature regime, the dynamics of the liquid start to be dominated by the landscape and the system tends to spend increasing time within individual basins corresponding to the IS.  Consequently,  an increased correlation may be expected between structural parameters for the liquid and the corresponding IS. To verify this, we inspect the relation between the values of our observables before and after quenching. In  Fig.~\ref{fig:2} a) and b) we show the $2d$ density plot of the $S_2$  and $\ntet$ values at temperature $T=0.7$. We can observe a clear correlation between the observables in the liquid state and the IS. Note that in Fig.~\ref{fig:2} b) there is an increase of correlation at a value of $\ntet=20$, corresponding to particles located in the center of icosahedral clusters. {Hence, we can deduce that some specific structural motifs remain unchanged when liquid configurations are mapped to the IS.}\\
Clearly, this correlation might be temperature dependent. In order to have a broader picture of the correlation between the two {sets of configurations (liquid and IS)} at different temperatures, we calculate Spearman's rank order correlation  \cite{spearman1987proof} between the values of $S_2$ and $\ntet$ per particles in the liquid and {the values in the corresponding IS}. The results presented in Fig.~\ref{fig:2} c) show that the correlation is stronger at lower temperatures, reaching a correlation of approximately 0.6 at $T=0.7$ for both $S_2$ and $\ntet$, demonstrating that the two-body entropy and tetrahedrality of the inherent structures are largely governed by the structure of the unquenched liquid as we observed in Fig.~\ref{fig:2} a) and b). As expected, above the onset temperature,  the correlation becomes negligible. This indicates that the local structure of a high temperature liquid is mostly insensitive to the underlying IS.\\
\subsection{Deformed glassy structures}
Now that we have characterized the static structure of our system using these descriptors, we can ask ourselves how these quantities evolve under cyclical shear deformation. To this end, we take the IS from the equilibrated configurations at a low and a high temperature of $T=0.7$ and $T=1.5$ and cyclically shear them with a range of strain amplitudes $\gamma_{max}$.  In particular, we distinguish two regimes of interest:\\
\textbf{I.} \textit{Below yielding}  ($\gamma_{max}< \gamma_y$): For small amplitudes of shear below yielding, the systems can anneal and, in the steady state, the system can visit lower energy minimum basins. In our system from $\gamma_{max}=0.02$ to $\gamma_{max}=0.06$, we have indeed  more annealed absorbing states: the steady-state energy is lower than the one of the initial configuration, in agreement with what has been observed in the literature~\cite{leishangthem2017yielding, bhaumik2019role} . If the initial configuration corresponds to a high-temperature liquid, the annealing effect is more pronounced in comparison to low-temperature initial configurations~\cite{bhaumik2019role}. In the steady state, the local plastic rearrangements during a cycle of deformation repeat from one cycle to the next~\cite{fiocco2014encoding}. Therefore, from one cycle to another, there is no appreciable net displacement when considering the configurations stroboscopically.\\
\textbf{II.} \textit{Above yielding} ($\gamma_{max} > \gamma_y$): In this case, the system becomes diffusive. In the steady state, the plastic rearrangements during a cycle lead to a net displacement of particles.
For a large enough system, like in our case, $N=64000$, it is possible to observe the formation of shear bands of highly diffusive particles \cite{fiocco2013oscillatory,parmar2019strain,priezjev2020shear}. In the case of $XZ$ shearing, the shear band consists of a slab of particles parallel either to the $YZ$ or the $XY$ plane. These particles on average present net displacements, between two consecutive cycles, larger than the rest of the particles. Interestingly, the region inside the shear band is characterized by a lower local density and a higher local energy \cite{parmar2019strain}.

\begin{figure}[!htb]
\begin{center}
  \includegraphics[width=1\linewidth]{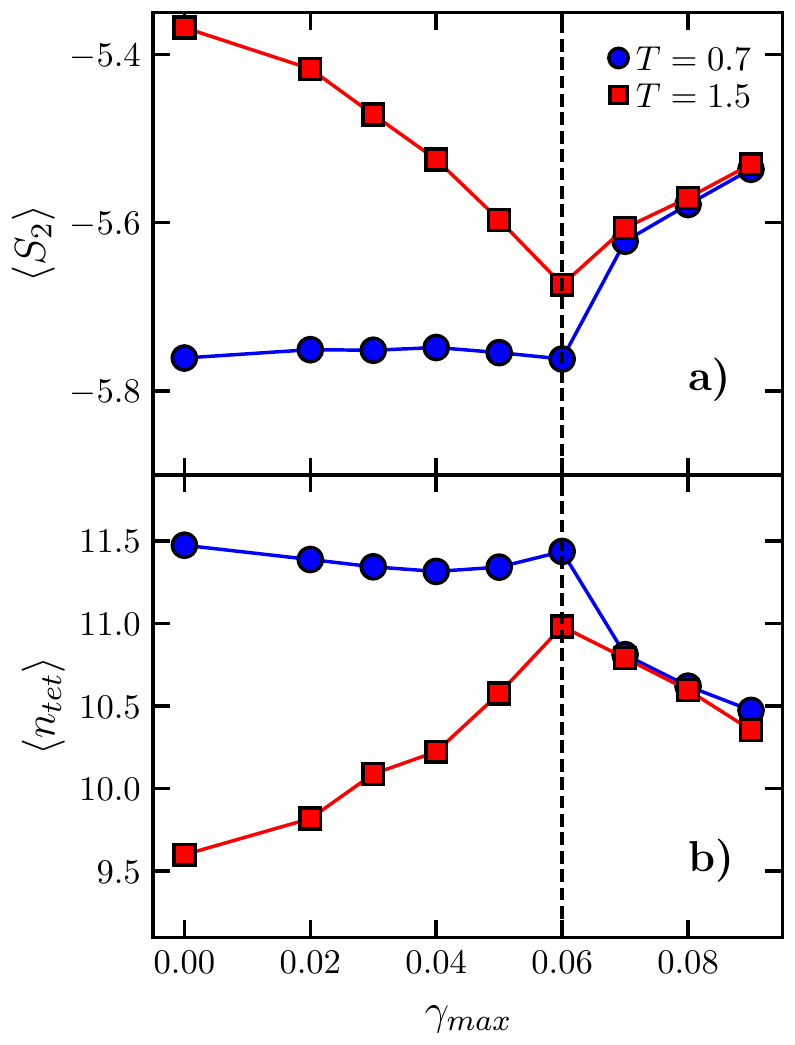}
  \caption{
  Behavior of a) $\sav$ and b) $\ntetav$ in the steady state as a function of $\gamma_{max}$, for cyclically sheared inherent states of equilibrium configurations obtained at $T = 0.7$ (blue) and $T = 1.5$ (red). The vertical dashed line shows the yielding amplitude $\gamma_y$. }
  \label{fig:3}
\end{center}
\end{figure}

Now that we have established the macroscopic effects of shear on our system, we examine the impact of the shearing amplitude on the local structure. To this end, we
first we compute $\sav$ and $\ntetav$  in the  steady state for strain amplitudes $\gamma_{max} \in [0.02,0.09]$, i.e. both below and above $\gamma_y\sim0.06$. In Fig.~\ref{fig:3}, we report the behaviour  of the mean value of these local descriptors as a function of $\gamma_{max}$, for temperatures $T=1.5$ and $T=0.7$. Note that $\gamma_{max}=0.0$ is actually the value for the initial IS configuration.  The high temperature glass in the range $\gamma_{max}\in[0.0,0.06]$ displays noticeable annealing and, as a consequence, an increase in  local order reflected by a lower $\sav$ and a higher $\ntetav$. As we cross $\gamma_y$, the diffusive state is reached,  and the average local order decreases significantly. This behaviour implies that remarkably, at yielding, a maximum local order is attained. Moreover,  the transition across $\gamma_y$ 
manifests in a cusp-like variation for both quantities.
For $T=0.7$, no strong annealing is observed below yielding and  $\sav$ and $\ntetav$ remain almost constant. Upon crossing $\gamma_y$,  the two observables  display a discontinuous jump.
Above yielding, the structural observables are the same for both temperatures, as has previously been observed for the energy~\cite{fiocco2013oscillatory}.
The cusp for the high temperature and the jump for the low one is compatible with the existence of a critical temperature that divides two different regimes or yielding behaviour. ~\cite{bhaumik2019role}.


To examine the effects of these changes in local structure on the dynamics of individual particles, we need to quantify how much a given particle moves during  a displacement cycle. When a system is sheared, the motion of the particles can be separated into affine and non-affine displacements~\cite{priezjev2016nonaffine}. One way of detecting and characterizing the local re-arrangements that the particles undergo during the cyclic deformations is through their non-affine displacement $D^2_{i}$ as introduced in Ref.~\onlinecite{falk1998dsqmin}. We employ a modified version of  $D^2_{i}$ where we calculate the non-affine displacement as a function of the accumulated deformation $\gamma$ instead of time: 

\begin{align}
    D^2_{i}(\gamma)=\frac{1}{N_i} \sum_{j=0}^{N_i} [(\mathbf{r}_j(\gamma)-\mathbf{r}_i(\gamma))-\Gamma(\mathbf{r}_j(0)-\mathbf{r}_i(0))]^2,
    \label{eqnDsq}
\end{align}
where $\mathbf{r}_{i}(0)$ is the position of particle $i$ at the beginning of a deformation cycle and $\mathbf{r}_{i}(\gamma)$ is its position in the deformed box at a total accumulated deformation of $\gamma$. 
The sum in equation \ref{eqnDsq} is over the $N_i$ nearest neighbors of particle $i$, taken to be the neighbors within a cut off $r_{cut}=1.4\sigma_{ij}$, which corresponds to the first minimum of $g(r)$ of the full system. The matrix $\Gamma$ is such that it minimizes the non-affine displacement giving a measure of local strain \cite{falk1998dsqmin}. This definition best maps deviation from affine displacement at a local level \cite{priezjev2016nonaffine}. 
In the case of a cyclic deformation below yielding, the system returns back to the same configuration each cycle once the steady state has been reached. Consequently,  if we consider configurations at the beginning ($\gamma=0$) and at the end  (after  $4\gamma_{max}/d\gamma$ deformation steps) of the cycle, we would observe $D^2_{min}(4 \gamma_{max}) \approx 0$. However,  we are interested in identifying which particles went through maximum re-arrangements during the deformation, regardless of reversibility. Therefore, we recorded the maximum value $\max(D^2_{min})$ for each particle in a cycle of deformation. 

\begin{figure}
\begin{center}
  \includegraphics[width=1\linewidth]{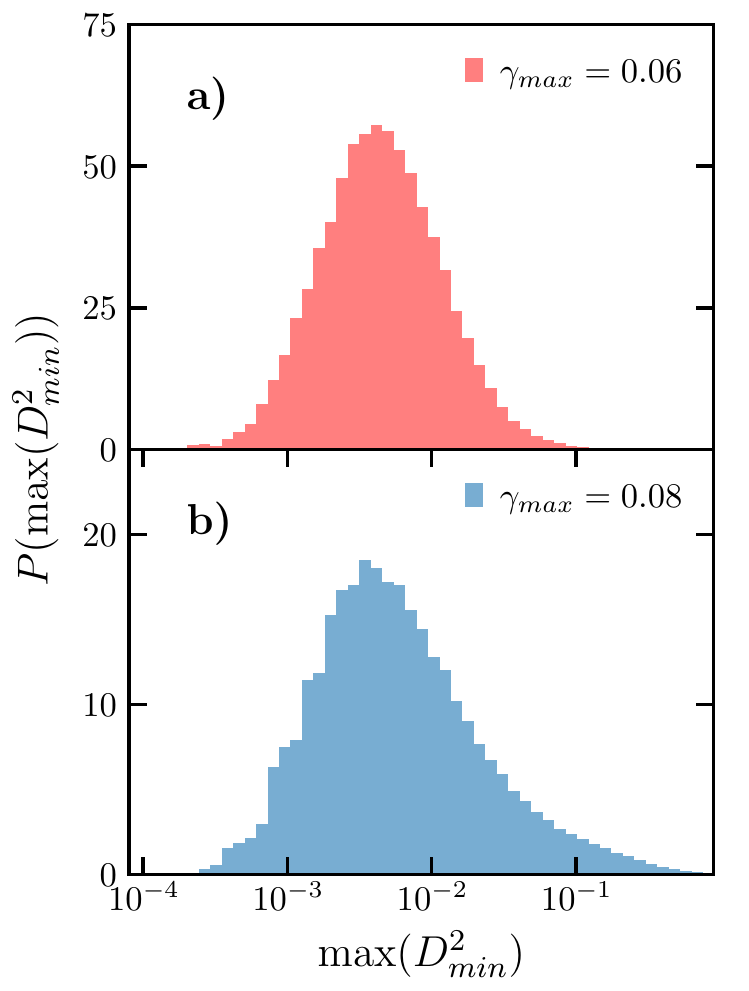}
  \caption{Distribution of $\max(D^2_{min})$ in the steady state at shear amplitude a) below yielding at $\gamma_{max}=0.06 < \gamma_y$ and b) above yielding at $\gamma_{max}=0.08 > \gamma_y$. Above yielding we have longer tail. In both cases, the temperature of the initial configuration was $T=0.7$. }
  \label{fig:4}
\end{center}
\end{figure}

In Fig.~\ref{fig:4}, we plot the distribution of  $\max(D^2_{min})$ for two strain amplitudes below and above yielding.  It is evident that, above yielding,  a  long tail is observed  in the distribution.  This tail is due to the particles in the shear band that present the maximum degree of plastic rearrangement. 

The calculation of $\max(D^2_{min})$  enables us to identify particles with high displacements, corresponding to those undergoing plastic rearrangements. At this stage it is natural to ask, is there any structural {difference} between the particles that undergo strong rearrangements and the particles that are immobile?
In other words, at the end of a cycle,  are particles with high and low $\max(D^2_{min})$ values characterized by a different structures at the beginning of a cycle? These are the questions that we will address in the remaining part of the paper.

\begin{figure}[!htb]
\begin{center}
  \includegraphics[width=1\linewidth]{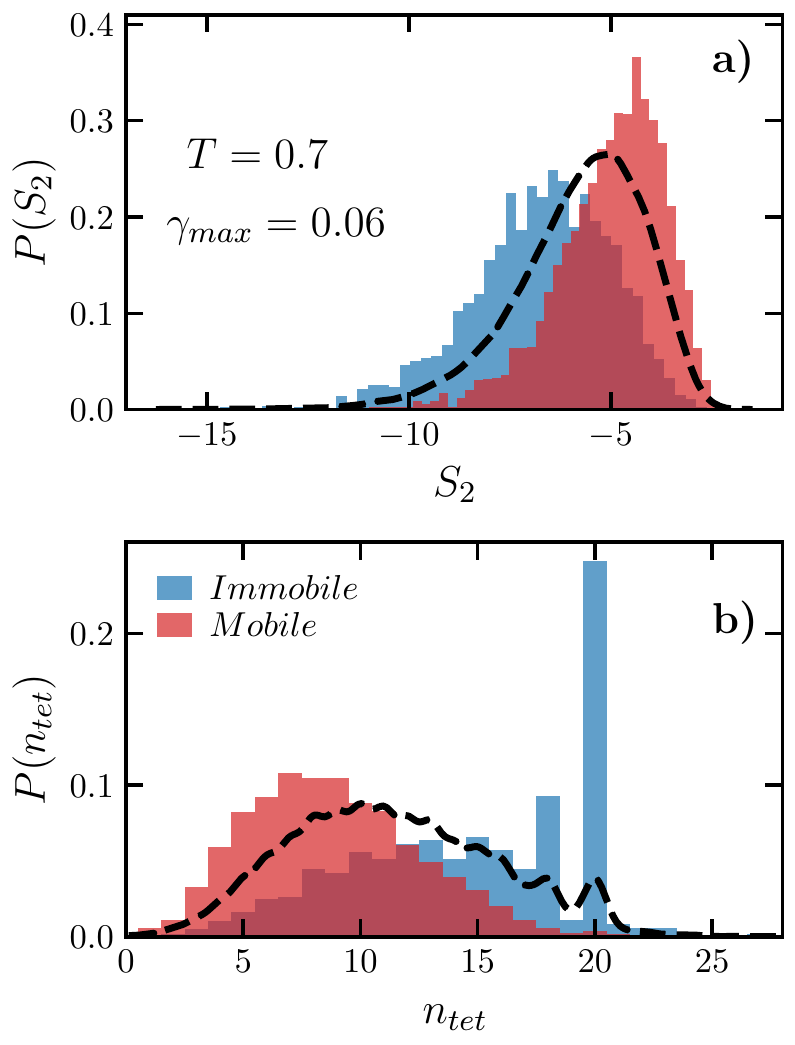}
  \caption{Distributions of a) $S_2$ and b) $n_{tet}$ for unsheared configurations in the steady state at temperature $T=0.7$ and shear amplitude $\gamma_{max}=0.06$. The black line corresponding to the overall distribution. Distributions are shown for immobile (blue) and mobile (red) particles, corresponding to the lowest and highest 5\% of displacements $\max(D^2_{min})$ during a shear cycle, respectively. }
  \label{fig:5}
\end{center}
\end{figure}

\begin{center}
 \begin{figure*}
  \includegraphics[width=1\linewidth]{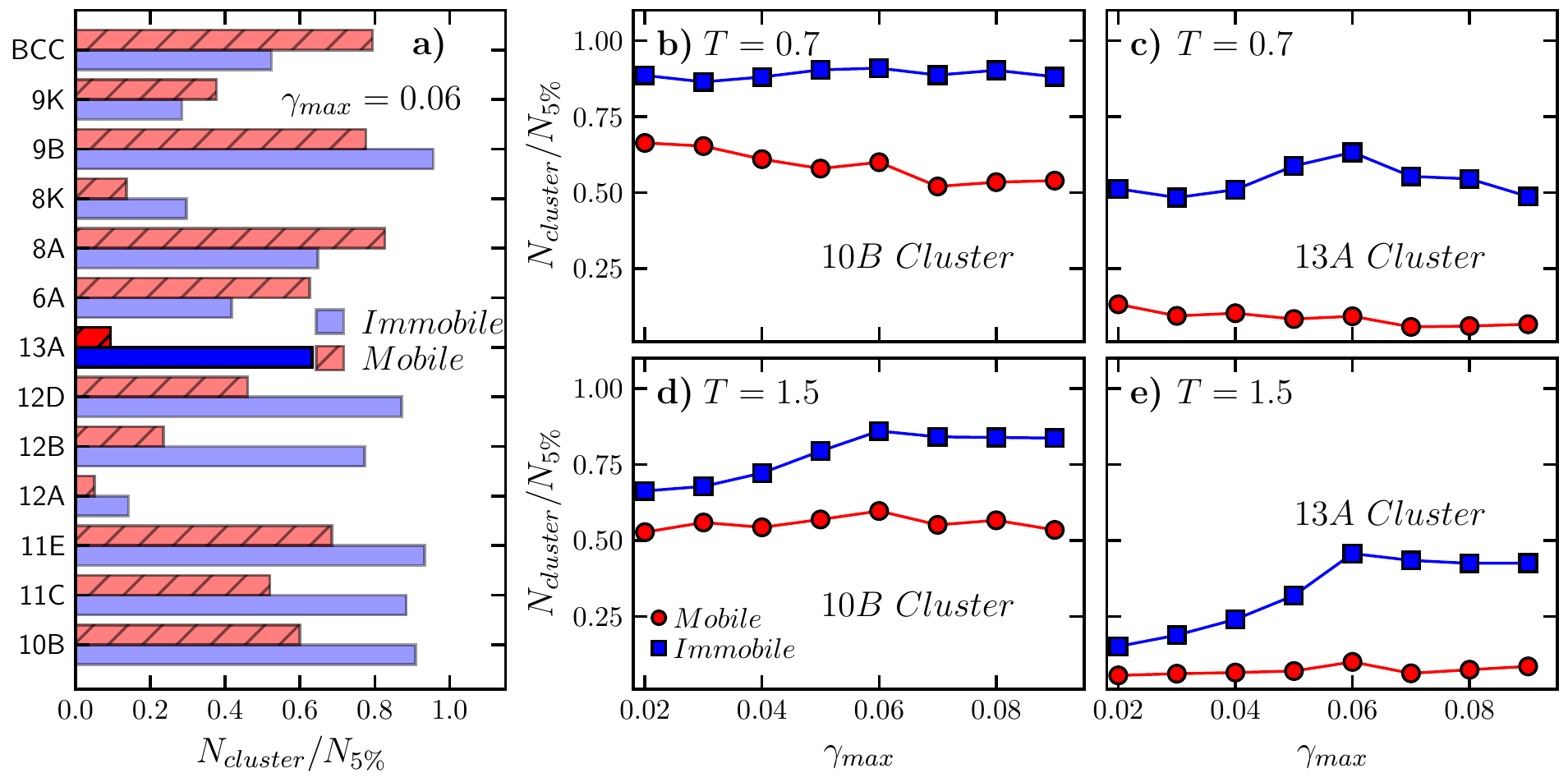}
  \caption{TCC analysis for mobile (red) and immobile (blue) particles. In a) for $T=0.7$ and $\gamma_{max}=0.06$ we report the fraction of these mobile and immobile particles in a particular type of cluster, where $13A$ is the icosahedral cluster. Next we plot these fractions against $\gamma_{max}$ for two particular classes b) $10B$ (defective icosahedral cluster) and c) $13A$ (icosahedral cluster) at a temperature of $T=0.7$ and the corresponding d) $10B$ and e) $13A$ at $T=1.5$. }
  \label{fig:7}
 \end{figure*}
\end{center}

To answer these questions, we turn our attention again  to the  structural local  descriptors $S_2$ and $n_{tet}$. In the steady state, we compute both descriptors for each particle in the stroboscopic configurations at a given strain amplitude. Subsequently, we subject the system to a complete deformation cycle during which we compute the values of $\max(D^2_{min})$ {for each} particle. As in previous studies on dynamical heterogeneities~\cite{kob1997dynamical},  we define a threshold based on the displacement of the individual particles  that allows  us to classify the particles as \textit{immobile} or \textit{mobile} if their $\max(D^2_{min})$ values are among the smallest or the largest respectively. In particular, we choose the top $5\%$ as mobile particles and bottom $5\%$ for the immobile ones. The resulting distributions are shown  in Fig.~\ref{fig:5}. Here,  we present  the case of $T=0.7$ and $\gamma_{max}=0.06$. However, the trend is similar in all the state points we investigated. From $P(S_2)$, we see that immobile particles have significantly lower $\sav$ implying a stronger local  order in comparison to the mobile particles. This trend is confirmed by the behaviour of $P(n_{tet})$ which shows that immobile particles are much more likely to be found in highly tetrahedral environments. Interestingly, for the immobile particles, we observe a strong peak at $n_{tet}=18$ and $n_{tet}=20$, the first corresponding to clusters formed by rings of 5 particles leading to high tetrahedrality and the second to particles at the center of an icosahedral cluster. This peak is essentially absent in the mobile particles, indicating that icosahedral caging has a strong effect on the local rearrangements in this system.  The appearance of this cluster has been investigated on many occasions in supercooled liquid systems as they represent the fingerprint of geometrical frustration introduced by Frank\cite{frank1952supercooling} and have been found to be strongly linked to slow dynamics in a wide variety of systems~\cite{jonsson1988icosahedral,kondo1991icosahedral,steinhardt1981icosahedral, jonsson1988icosahedral,leocmach2012roles, hirata2013geometric, ding2014full, marin2019slowing, vasisht2020emergence}.
\begin{figure}[!htb]
\begin{center}
  \includegraphics[width=1\linewidth]{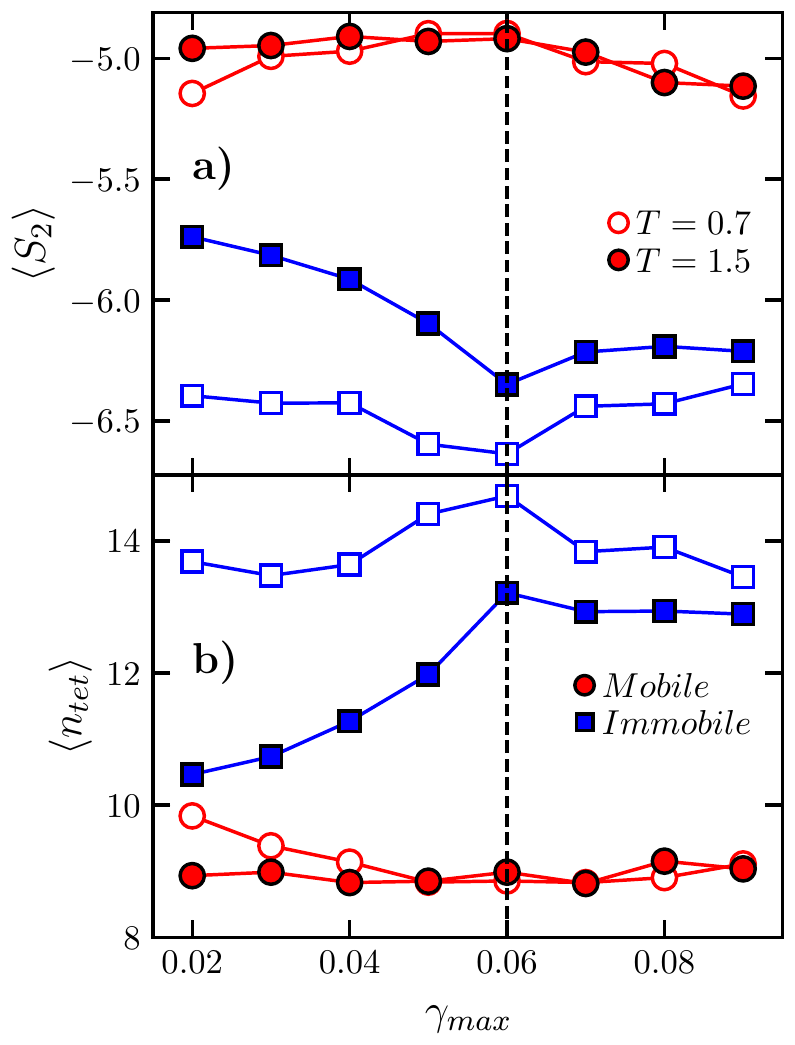}
  \caption{Behavior of a) $\sav$ b) $\ntet$ as a function of $\gamma_{max}$ in the steady state, for the $5\%$ most mobile (red circle) and $5\%$ most immobile particles (blue squares). Filled symbols correspond to $T=1.5$ and open symbols to $T=0.7$.}
  \label{fig:6}
\end{center}
\end{figure}
We now turn our attention to the average structural characteristics of the mobile and immobile particles.
In Fig.~\ref{fig:6},  we plot $\sav$ and $\ntetav$ for immobile and mobile particles as a function of strain amplitude $\gamma_{max}$ and for both high ($T=1.5$)  and low ($T=0.7$) temperatures. 
For the immobile particles, below yielding,  $\ntetav$ increases and $\sav$ decreases. These trends are more pronounced for $T=1.5$ and can be ascribed to the stronger tendency to anneal and the consequent increase of local order, as we observed in Fig.~\ref{fig:3}. Above yielding, the immobile particles are {present}
outside the shear band. We find that upon crossing the yielding amplitude ($\gamma_{max}>0.06$), both $\sav$ and $\ntetav$ flattens out for the high temperature case.
For 
the mobile particles, as expected, the values of $\sav$ and $\ntetav$  are consistent with a less structured local structure. However, we notice 
that their structural features change very little between the two temperatures. In fact,  both $\sav$ and $\ntetav$ remain almost constant below and above yielding. Although the origin of this effect is not clear, it suggests that the most mobile particles are characterized by a similar local structure, which is independent of temperature, shear amplitude, and even the presence of a shear band. {Equally interestingly, even though they do not participate in the shear bands in which strain is localized, the structure around immobile particles is significantly affected by shear band formation.}\\
So far, we have established that the particles characterised by different propensity to  re-arrange  are embedded, on average,  in different local environment  as reflected by the local descriptors considered here. However, we have not discussed the actual geometrical shape of the local cluster around  a particle and how it is connected with plastic rearrangement.  To this end,   we classify the local arrangement of particles by using the Topological Cluster Classification algorithm (TCC)\cite{malins2013identification}. This algorithm detects predefined specific local clusters of different sizes. The detection is based on the connections between nearest neighbors, found via a modified Voronoi construction. As a first step, simple small clusters from 3 up to 7 particles are identified. Subsequently, more complex clusters are found as combinations of the simple ones. Note that each particle can simultaneously be part of multiple clusters. For the $5\%$ most mobile and immobile particles we examine all possible clusters detected by TCC and we calculate the fraction of particles involved in a certain type cluster for each case. In Fig.~\ref{fig:8} a) we show the results for mobile (red) and immobile (blue) particles for the case of $T=0.7$ and $\gamma_{max}=0.06$. Here, we report only the cluster types that show significant differences between the two communities. In particular, we highlight the $13A$  cluster corresponding to an icosahedral cluster for which the largest difference between mobile and immobile particles is noticed. Similarly, another cluster that presents large differences, but less pronounced than the $13A$ case, is the $10B$ cluster which corresponds to a defective icosahedral cluster. It appears that these two clusters play an important role in determining whether a given particle will rearrange during a shear cycle. Note that the two of them in particular can be decomposed in several tetrahedral clusters characterised  a particularly high value of $\ntet$ value. For example, for a particle located in the center of an icosahedral cluster, its $\ntet=20$ as each of the faces of the icosahedral form a tetrahedral cluster with the central particle. In the following, we will focus mainly on these two clusters.\\
In particular, in Fig.~\ref{fig:7}(b-e),  we show the fraction of particles involved  in the $10B$ and $13A$ cluster for each of the communities as a function of $\gamma_{max}$ as they present the largest difference.  
For the immobile particles at $T=0.7$, the population of $10B$ clusters is essentially constant, while the icosahedral cluster displays a noticeable maximum at the yielding transition. For $T=1.5$, below yielding, there is a more pronounced effect as the fraction of immobile particles involved into these clusters grows with the amplitude $\gamma_{max}$. Above yielding, this effect flattens out as no noticeable increase is observed.
For mobile particles at both temperatures, it appears that the variation in the population of the two clusters is negligible. Similar results are found for other TCC clusters that differ in prevalence between the immobile and mobile particles, such as the ones shown in Fig.~\ref{fig:8} a). Essentially, there are no appreciable changes in the structure of the mobile particles with different amplitudes. In contrast, the immobile ones show significantly different behavior across the yielding transition.

These trends are consistent with the variation of $\sav$ and $\ntetav$ in Fig.~\ref{fig:6} and seems to confirm that the local environment in which the particles are embedded plays an important role. In particular, it is interesting to note that despite the immobile particles displaying a limited activity in terms of plastic rearrangements, there is a more remarkable variation of their local structure upon the change of amplitude of the deformation. 


From Fig.~\ref{fig:5} b) we see that a large fraction of $5\%$ most immobile particles are characterized by $n_{tet}=20$, which is the number of tetrahedra that one would expect in an icosahedral cluster. We confirm this idea by inspecting Fig.~\ref{fig:7} c) and e), where we  notice that the $5\%$ most mobile and $5\%$ least mobile particles on average show significant difference regarding to the icosahedral ordering. 

\subsection{Glassy structure and the shear band}

As discussed in the introduction, above yielding, the dynamics become heterogeneous and localized in space and, as a consequence, shear band formation is expected. There are numerous hints that the properties of the structure inside the shear band is different from what is observed outside. It has been observed, for a similar model to the one discussed here, that the density inside the shear band is lower~\cite{parmar2019strain} and that it may present a certain degree of hyperuniformity~\cite{Mitra_2021}.  The question we want to answer next is whether its structure is different and if $\sav$ and $\ntetav$ capture the  differences inside and outside the shear band.

\begin{center}
 \begin{figure}[!htb]
  \includegraphics[width=1\linewidth]{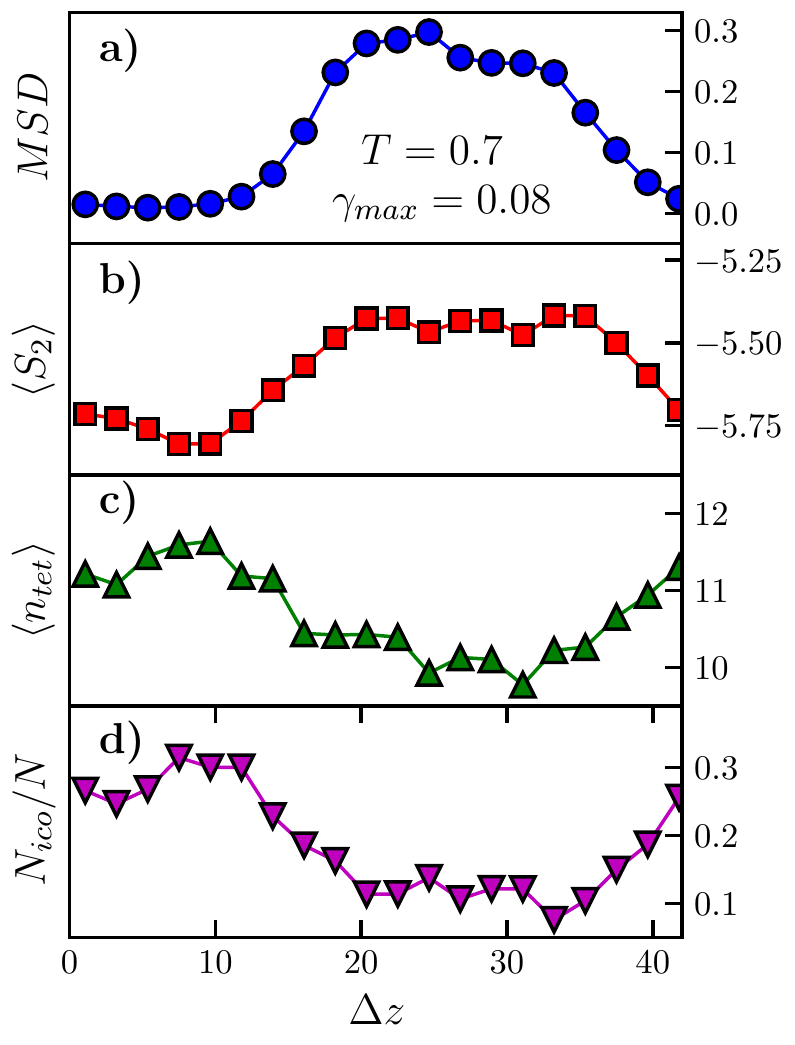}
  \caption{Shear band captured by the different properties a) mean-squared displacement, b) $S_2$, c) $\ntet$ and d) fraction of particles involved in icosahedral clusters $N_{Ico}/N$ as a function of the perpendicular axis to the shear band. All corresponding to sheared system at a $\gamma_{max}=0.08$ at a temperature $T=0.7$.  }
  \label{fig:8}
 \end{figure}
\end{center}

To detect the shear band, we bin the system along the axis perpendicular to the plane of the shear band. In this case this corresponds to the $z$-axis. Subsequently, we compute the mean square displacement (MSD) of the particles between the $n^{th}$ and $(n+1)^{th}$ cycle  as a function of $z$ \cite{parmar2019strain}. The position of the  shear band can be clearly identified as the region where the particles have a larger MSD. The result, for the case of $T=0.7$, is presented in  Fig.~\ref{fig:8} a). Along the same axis, we measure $\sav$ and $\ntetav$ and show the results in Fig.~\ref{fig:8} b) and c) respectively. Both these structural features indicate strong structural differences between the portions of the system inside and outside of the shear band. On  average, outside the shear band we observe higher structural order.  
In Fig.~\ref{fig:8} d),  we show the distribution of the icosahedral clusters along the direction perpendicular to the shear band.  We find a clear localization, where $30\%$ of the particles outside the shear band are involved in an icosahedral cluster, whereas inside the shear band this fraction is {much lower.}
This observation suggests that the lower density that has been observed inside the shear band~\cite{parmar2019strain} is related to a pronounced structural signature. In particular, the systems prefers to localize a great number of icosahedral structures, that are know to be more energetically favourable, outside the more mobile region. This localization is not found below yielding, where the icosahedral clusters are homogeneously distributed inside the simulation box.

\section{Conclusions}

Predicting  plastic re-arrangements based on  local structures information  is a challenging domain of research (see Ref.~\onlinecite{richard2020predicting} and reference therein) that has attracted a lot of interest in recent years~\cite{xu2021atomic, fan2021predicting, shimada2021spatial, kriuchevksyi2021predicting}. In this paper, we have studied the local structural rearrangements of  particles undergoing cyclic shear deformation in a model glass former  focusing on local tetrahedrality $\ntet$ that we introduced previously~\cite{marin2020tetrahedrality} and the well known two-body excess entropy $S_2$. High values of $\ntet$ and low values of $S_2$ indicate a higher degree of {local order.}
In particular, we consider two temperatures, above and below the onset temperatures, for which different levels of initial average order is observed.  The {inherent structure} configurations were sheared following  a cyclic athermal quasistatic deformation protocol  that revealed the typical phenomenology of the weakly (high temperature) and deeply (low temperature) annealed glass~\cite{bhaumik2019role}. We then investigated the links between local structure and local rearrangements at different strain amplitudes $\gmax$ below and above yielding.\\
Below yielding, the system anneals, resulting in an increase of local structural order until the system reaches a periodic state and no longer diffuses. In contrast, above yielding both the low-temperature and high-temperature glasses become diffusive and exhibit shear banding. Moreover, above yielding, we found that both systems present on average the same values of $\sav$ and $\ntetav$ irrespective of its preparation history. This is compatible with the phenomenology observed before in several different systems~\cite{fiocco2013oscillatory,regev2013onset,keim2014mechanical,priezjev2016reversible,leishangthem2017yielding,ozawa2018random,bhaumik2019role}.\\
During the deformation cycle in steady states below yielding,  particles rearrange reversibly, meaning that at the end of the cycle they come back to their original positions. In contrast, above yielding the rearrangements are irreversible and particle,  inside shear band,   display  diffusive motion. 
By investigating the local structure around the 5\% most immobile and most mobile particles during a given shear cycle, we found that the immobile and mobile particles have  on average different local order at the beginning of the cycle. In particular, particles with lower $\ntet$ and higher $S_2$ are more likely to rearrange, an observation which holds for all investigated $\gmax$ and temperatures. Moreover, the two temperatures explored here correspond to glass that would have a brittle (low temperature) and  ductile (high temperature) mechanical response~\cite{ozawa2018random, bhaumik2019role}
corresponding to different levels of annealing.
Interestingly, for the low temperature, the average local structures do not display significant evolution below yielding. However, if we consider only the immobile particles, we observe a clear evolution, for example, in the number of tetrahedra. This indicates that the main effect of annealing below yielding is the emergence of a population of particles with high local order, which are highly stable against further rearrangement. \\
In line with these results, we closely investigated the structure inside and outside shear band. We showed that both of our local structure descriptors are spatially correlated to such shear band. In particular, outside it, the local structure is more ordered presenting higher values of $\ntet$ and lower of $S_2$. Interestingly, we found that outside the shear band more than $30\%$ of the particles are involved in icosahedral clusters whereas inside it this percentage is essentially negligible. In other words, the population of highly ordered particles that emerges upon annealing persists even beyond yielding, but remains localized outside the shear band, where they may have a stabilizing influence on their surroundings. {Our results confirm that, as for the case of continuous shearing~\cite{richard2020predicting}, certain aspects of the  response to cyclic shear deformation can be correlated to  structural features}.

\begin{acknowledgements}
The authors would like to thank Davide Fiocco for the original production of the potential energy cartoons in Fig.~\ref{fig:0}. S.M., S.S and G.F.  gratefully acknowledge IFCPAR/CEFIPRA for support through project no. 5704-1. S.M.-A. acknowledges support from the Consejo Nacional de Ciencia y Tecnología (CONACyT scholarship No. 340015/471710). SS acknowledges support through the JC Bose Fellowship  (JBR/2020/000015) SERB, DST (India).
\end{acknowledgements}

\bibliographystyle{apsrev4-1}
\bibliography{references}

\end{document}